\documentclass[pre,reprint,superscriptaddress,showpacs,groupedaddress]{revtex4-1}
\usepackage{graphicx,amsfonts,amsmath,amssymb,color}
\usepackage[english]{babel}
\usepackage{natbib}

\newcommand{\e}{\varepsilon}

\newcommand{\V}[1]{\mathbf{#1}}

\newcommand{\intl}[3]{\int\limits_{#1}^{#2}\mathrm{d}#3\,}
\newcommand{\bracket}[1]{\left(#1\right)}

\newcommand{\grb}[1]{\mbox{\boldmath $#1$}}

\newcommand{\eq}[1]{$\mathrm{Eq.}$~\eqref{#1}}
\newcommand{\eqs}[1]{$\mathrm{Eqs.}$~\eqref{#1}}
\newcommand{\figref}[1]{$\mathrm{Fig.}$~\ref{#1}}
\newcommand{\secref}[1]{$\mathrm{Sec.}$~\ref{#1}}

\DeclareMathOperator{\asinh}{asinh}

\newcommand{\av}[1]{\left\langle #1 \right\rangle}

\begin{document}

\title{Front propagation in channels with spatially modulated cross-section}

\author{S. Martens}
\email{steffen.martens@tu-berlin.de}
\author{J. L\"ober}
\author{H. Engel}
\affiliation{Institut f\"ur Theoretische Physik, Hardenbergstra\ss e 36, EW 7-1, Technische Universit\"at Berlin, 10623 Berlin, Germany}

\begin{abstract}

The problem of front propagation in a three-dimensional channel with spatially varying cross-section is reduced to an equivalent reaction-diffusion-advection equation with boundary-induced advection term. Treating the advection term as a weak perturbation, an equation of motion for the front position is derived. We analyze channels whose cross-sections vary periodically with $L$ along the propagation direction of the front. Taking the Schl\"{o}gl model as representative example, we calculate analytically the nonlinear dependence of the front velocity on the ratio $L/l$ where $l$ denotes the intrinsic front width. Our analytical results agree well with the results obtained by numerical simulations. In particular, the peculiarity of boundary-induced propagation failure for a finite range of $L/l$ values is predicted by analytical calculations. Lastly, we demonstrate that the front velocity is determined by the suppressed diffusivity of the reactants for $L \ll l$.
%
\end{abstract}

\maketitle

\section{Introduction}

\noindent Propagating fronts are building blocks of traveling wave (TW) activity in dissipative spatially
extended systems. Like excitation pulses, periodic pulse trains, spiral and scroll waves, they are examples of nonlinear waves and represent fascinating self-organized spatio-temporal patterns in nonequilibrium macroscopic systems. Traveling waves have been observed in many physical \cite{Hohenberg1993}, biological \cite{Murray2003,Bressloff2013,Newman1997}, and chemical systems \cite{Kuramoto2003,Kapral1995}. Prominent examples of front propagation include catalytic oxidation of carbon monoxide (CO) on platinum single crystal surfaces \cite{Baer1992,*Baer1995,Voroney1996,Shvartsman1999,Rotermund2009}, arrays of coupled chemical reactors \cite{Laplante1992}, and nematic liquid crystals \cite{Haudin2009}.

Often, the medium that supports front propagation exhibits a complex shape and/or its size is limited like in biological cells \cite{Petty2008}, nanoporous media \cite{Atis2013}, or zeolites \cite{Beerdsen2006}. In such system the interaction of the reactants with the boundaries of the medium leads to non-intuitive confinement effects \cite{Toth1994,Garcia1999,*Garcia2001}. For example, phase separation in porous materials results in layering, freezing, wetting and other novel phase transitions not found in the bulk system \cite{Gelb1999}.
In particular, chemical reactions \cite{Santamaria2013,Becker2014} as well as molecular diffusion \cite{Verkman2002,Corma1997,Keyer2014,*Keyer2014arxiv} depend strongly on the shape of the domain. Recent studies on the fundamental problem of particle transport through micro-domains exhibiting obstacles and/or small openings showed that the shape of these confinements (periodicity, size of the connecting openings) regulates the dynamics of diffusing particles leading to transport properties which significantly differ from free Brownian motion \cite{Reguera2006,Burada2008,Burada2009_CPC,Martens2013,Becker2013}.

Even in systems ranging from micro- to the macroscale there is a ongoing interest in studies of nonlinear wave propagation under spatially confined conditions. Important issues investigated in this context are the dependence of front reflection on the geometry size in the catalytic CO oxidation on platinum surfaces \cite{Haas1995},
reaction fronts in Poiseuille \cite{Edwards2002} or shear flows \cite{Vasquez2004}, three-dimensional ($3$D) traveling waves in the human heart \cite{Fenton2011}, to name a few. Some parts of the human heart tissue, especially at the ventricles, are thick enough to support not only spiral waves, but also $3$D vortex structures, for example scroll waves and scroll rings. Hence, detailed knowledge about the interaction of these self-organized spatio-temporal patterns with boundaries \cite{Azhand2014arxiv,Totz2014arxiv} might be important for the understanding of possible mechanisms responsible for atrial tachycardia. In particular, there is experimental evidence that spatial variations of the heart wall's thickness play a significant role in atrial fibrillation \cite{yamazaki2012}.

Nowadays, well-established techniques like micro\-lithography enable to design the shape of reactive domains as well as to prescribe the boundary conditions \cite{Rotermund2009}. This provides an efficient method to study experimentally the impact of confinement on front propagation and to control, respectively, optimize the local dynamics of catalytic reactions.

In this paper, we address the problem of traveling front (TF) propagation through a $3$D channel with periodically modulated cross-section. Aiming at deriving an equation of motion for the front position in corrugated channels, we apply asymptotic perturbation analysis in a geometric parameter \cite{Martens2011,*Martens2011b,Martens2013} and projection techniques \cite{Schimansky1983,*Engel1985,Biktashev2003,*Biktashev2009,Loeber2012PRE,Loeber2014PRL} to the problem. Our goal is to analyze how spatial variations of the channel's cross-section affect front propagation and, in particular, to determine the dependence of the propagation velocity on the characteristic length scales in the system. Furthermore, we focus on boundary-induced phenomena such as propagation failure and effective diffusivity.

The paper is organized as follows: In \secref{sec:model}, we formulate the reaction-diffusion (RD) equation in a $3$D channel with spatially modulated cross-section. In \secref{subsec:asym} and \ref{subsec:proj}, 
we derive an equation of motion for the traveling front using multiple scale analysis. Additionally, we obtain analytical expressions for the average front velocity. In \secref{sec:numerics}, we compare our theoretical results with numerical simulations for the one component Schl\"ogl model. An analytical estimate for the interval of propagation failure is presented in \secref{subsec:eikonal}. In \secref{sec:diff}, we demonstrate that the front velocity is determined by the suppressed diffusivity of the reactants if the intrinsic width of the front is much larger than the spatial variation of the channel's cross-section. Finally, we conclude our results in \secref{sec:conc}.

\section{Statement of the problem} \label{sec:model}
\begin{figure}[t]
  \centering
  \includegraphics[width=0.95\linewidth]{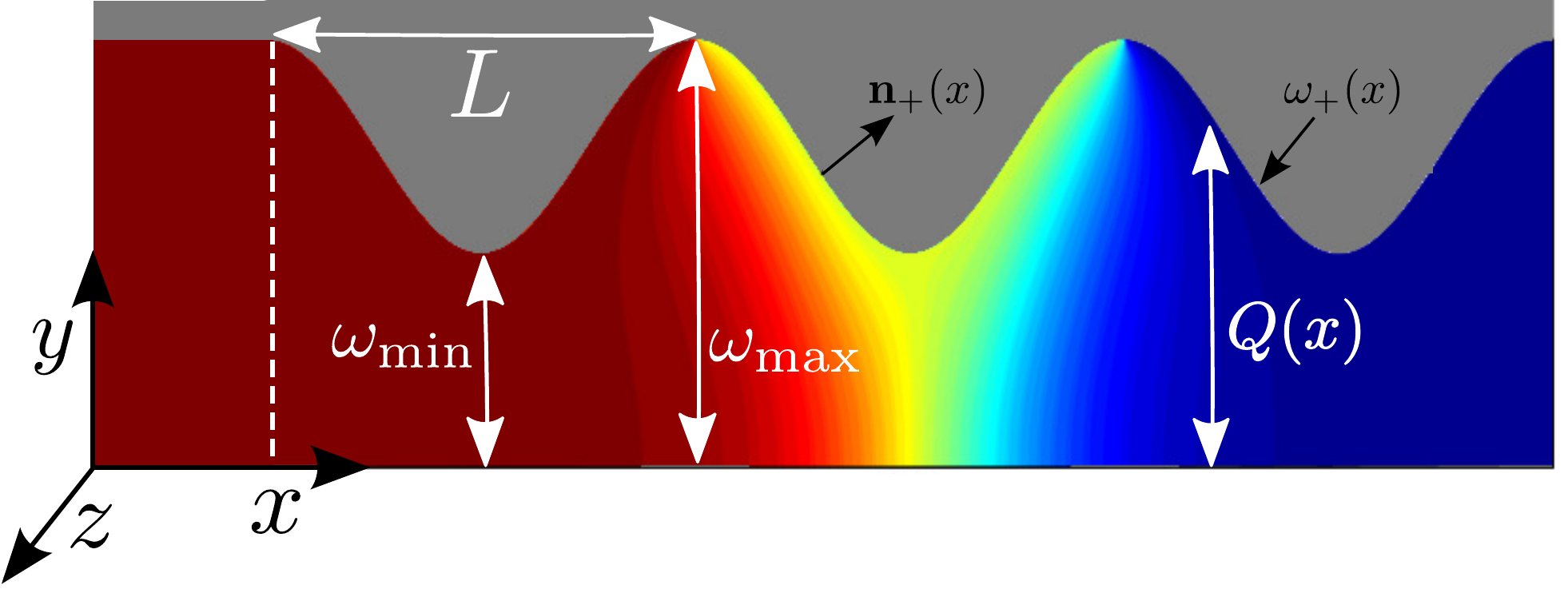}
  \caption{(Color online) Plan view on a segment of the channel with spatially modulated cross-section $Q(x)$. The channel is confined by a periodically modulated boundary at $y=\omega_+(x)$, with spatial period $L$, a flat boundary at $y=\omega_-(x)=0$, and by plane walls placed at $z = 0$ and $z = H$;  $H$ represents the channel height. The bottleneck size and the maximum channel width are $\omega_\mathrm{min}$ and $\omega_\mathrm{max}$, respectively. 
  The dashed line represents the begin of the boundary corrugation in the numerical simulations, cf. \eq{eq:wall}. Superimposed is the concentration field $u(\V{r},t)$ of a front traveling from left to right.}
  \label{fig:Fig1}
\end{figure}

\noindent We consider a RD system for the vector of $n$ concentration fields $\V{u}=\V{u}(\V{r},t)=\bracket{u^1,\ldots,u^n}^T$ whose spatial and temporal evolution is modeled by a reaction-diffusion equation
\begin{align}
 \partial_t \V{u}=\,\mathbb{D}\Delta\V{u}+\V{R}\bracket{\V{u}} \label{eq:RD}
\end{align}
in a channel filled with a excitable medium. Here, $\V{r}=\bracket{x,y,z}^T$ is the position vector, $\mathbb{D}=\mathrm{diag}(D_1,\ldots,D_n)$ represents the diagonal matrix of constant diffusion coefficients, $\Delta$ denotes the Laplacian operator, and $\V{R}\bracket{\V{u}}$ represents the nonlinear reaction kinetics. The medium filling the channel is assumed to be uniform, isotropic, and infinitely extended in the $x$-direction. In transverse directions, the channel is confined by periodically modulated walls at $y=\omega_\pm(x)$, with spatial period $L$ and plane walls placed at $z = 0$ and $z = H$. A sketch of the setup is depicted in \figref{fig:Fig1}. Because of the channel walls' impermeability with respect to diffusion the gradient of $\V{u}$ obeys no-flux boundary conditions (BCs), reading
\begin{align}
 \bracket{\nabla \V{u}\bracket{\V{r},t}}\cdot\V{n}\bracket{\V{r}}=0\,,\quad \forall \V{r} \in \mbox{channel wall}\,. \label{eq:bc}
\end{align}
$\V{n}(\V{r})$ denotes the outward-pointing normal vector at the channel walls, viz. $\V{n}_z=(0,0,\pm 1)^T$ at $z = 0, H$ and $\V{n}_\pm=(\mp \omega_\pm'(x),\pm1,0)^T$ at $y=\omega_\pm(x)$ with the prime denoting the differentiation with respect to $x$.

Passing to dimensionless quantities, all lengths are measured in units of the widest cross-section of the channel $\omega_\mathrm{max}$, yielding  $\V{r} \to \omega_\mathrm{max} \, \V{\overline{r}}$. 
The time is scaled in units of the inverse characteristic kinetic constant of the slowest reaction $t\to \overline{t}/k_\mathrm{reac}$. Thus, the re-scaled diffusion constants read $\mathbb{D} \to \omega_\mathrm{max}^2 k_\mathrm{reac} \overline{\mathbb{D}}$. For convenience, the channel height in units of $\omega_\mathrm{max}$ is set to unity $H=1$.

\subsection{Asymptotic perturbation analysis} \label{subsec:asym}
\noindent Below, we perform asymptotic perturbation analysis in the dimensionless geometric parameter
\begin{align} \label{eq:epsilon}
\e=(1 - \delta)/\overline{L} \ll 1,
\end{align}
which characterizes the deviation of corrugated $\omega_\pm(x)$ from flat boundaries, i.e., $\e=0$. $\delta$ denotes the ratio of the bottleneck width $\omega_\mathrm{min}$ to the maximal channel width $\omega_\mathrm{max}\equiv 1$, i.e., $\delta=\omega_\mathrm{min}/\omega_\mathrm{max}$. The choice for the expansion parameter $\e$ is motivated by previous studies on Brownian motion in corrugated channels \cite{Martens2011,*Martens2011b,Martens2013}. Upon re-scaling the transverse coordinate $y\to \e\,\overline{y}$, the profile functions become $\omega_\pm(\overline{x})\to \e\,h_\pm(\overline{x})$ and the outward-pointing normal vector at the perpendicular side-walls is given by $\V{n}_\pm= \bracket{ \mp \e h_\pm'(x), \pm 1,0}^T$. Thus, the no-flux BC \eq{eq:bc} transforms to
\begin{align}
 \bracket{\nabla \V{u}\bracket{\V{r},t}}\cdot\V{n}\bracket{\V{r}}= 0 =\,\mp \e^2 h_\pm'(\overline{x})\partial_{\overline{x}} \overline{\V{u}} \pm \partial_{\overline{y}}\overline{\V{u}}, \label{eq:bc_scal}
\end{align}
at $\overline{y}=\,h_\pm(\overline{x})$. In the following, we shall omit the overbar in our notation.

Expanding $\V{u}$ in a series in even powers of $\e$, we get $\V{u}(\V{r},t)=\V{u}_0(\V{r},t)+\e^2\,\V{u}_1(\V{r},t)+\mathcal{O}(\e^4)$. Substituting
this ansatz into \eq{eq:RD} and observing the no-flux BCs \eq{eq:bc_scal}, we obtain a hierarchic set of coupled partial differential equations (PDEs). In leading order, one has to solve $\mathbb{D}\partial_y^2 \V{u}_0=\,0$ supplemented with no-flux BCs, $0=\,\pm \partial_y \V{u}_0$ at $y=h_\pm(x)$ and $0=\,\pm \partial_z \V{u}_0$ at $z=0,H$.
It immediately follows that the concentration profiles $\V{u}_0(\V{r},t)=\V{g}(x,t)$ are flat in $y$ and $z$ direction. The unknown function $\V{g}(x,t)$ has to be determined from
the second order $\mathcal O(\e^2)$ balance given by
\begin{subequations}
 \begin{align}
 \partial_t \V{u}_0 =&\,\mathbb{D}\bracket{\partial_x^2 \V{u}_0 + \partial_y^2 \V{u}_1 +\partial_z^2 \V{u}_0}+\V{R}\bracket{\V{u}_0}.
 \intertext{Integrating the latter over the scaled cross-section $H \bracket{h_+(x)-h_-(x)}$ and taking into account the no-flux BCs}
 0=&\,\mp h_\pm'(x)\partial_x \V{u}_0 \pm \partial_y \V{u}_1,\, \forall y=h_\pm(x), 
\end{align}
\end{subequations}
one obtains
\begin{align} \label{eq:RDA}
 \partial_t \V{u}_0(x,t) =\,\mathbb{D}\partial_x^2 \V{u}_0+ \V{R}\bracket{\V{u}_0} + \mathbb{D} \frac{Q'(x)}{Q(x)}\partial_x \V{u}_0.
\end{align}
By applying asymptotic perturbation analysis in the small parameter $\e$, the Neumann boundary conditions on the reactants, \eq{eq:bc}, in the original $3$D channel with spatially varying cross-section translate into a one-dimensional reaction-diffusion-advection equation with a \textit{boundary-induced advection term}, \eq{eq:RDA}. The advective velocity field $\V{v}= Q'(x)/Q(x)\V{e}_x$ reflects the periodicity of the channel modulation $L$, $\V{v}(x+L)= \V{v}(x)$, and has zero mean, $\int_{0}^{L}\mathrm{d}x\,\V{v}(x)=\V{0}$. Referring to \eq{eq:RDA}, a front propagating from left to right, $\partial_x \V{u}_0 < 0$, becomes decelerated where the channel expands, $Q'(x)>0$, and accelerated if the channel contracts, $Q'(x)<0$, respectively. For systems where diffusion, advection, and reaction coefficients depend periodically on space and time it has been shown \cite{Xin1993} that the profile of a traveling front and its velocity change periodically in time -- the TF solutions are called \textit{
pulsating traveling fronts}. A lot of work have been done to proof the existence and stability of these pulsating TFs \cite{Xin2000,Berestycki2007,nadin2010}.

\subsection{Projection method -- multiple scale analysis} \label{subsec:proj}

\noindent Next, we derive the equation of motion for the TW's position in response to the boundary-induced advection term, $\mathbb{D}\, \bracket{\V{v}\cdot\nabla}\V{u}_0 \propto Q'(x)$, assuming weak spatial variations of the cross-section in propagation direction, i.e., $\mathrm{max}\bracket{|Q'(x)|}\simeq \e$. Following \cite{Loeber2014PRL}, we can treat the advection term as a weak perturbation to the $1$D RD system for the leading order $\V{u}_0$
\begin{align} \label{eq:RD1D}
 \partial_t \V{u}_0(x,t) =\,\mathbb{D}\partial_x^2 \V{u}_0+ \V{R}\bracket{\V{u}_0}.
\end{align}
We presume that the RD system \eq{eq:RD1D} possesses a stable TW solution $\V{U}_c$. This solution is stationary in frame of reference $\xi=x-c_0t$ co-moving with the velocity $c_0$
\begin{align}
 0=\,\mathbb{D}\partial_\xi^2\V{U}_c+c_0\partial_\xi \V{U}_c + \V{R}\bracket{\V{U}_c}.
\end{align}
The eigenvalues of the linear operator
\begin{align}
 \mathcal{L}=\,\mathbb{D}\partial_\xi^2+c_0\partial_\xi +\mathcal{D}\V{R}\bracket{\V{U}_c}
\end{align}
determine the stability of the TW, where $\mathcal{D}\V{R}\bracket{\V{U}_c}$
denotes the Jacobian matrix of $\V{R}$ evaluated at $\V{U}_c$.
Since we presume that $\V{U}_c(\xi)$ is stable, the eigenvalue of $\mathcal{L}$
with the largest real part is $\lambda_0=0$ and the Goldstone mode $\V{W}(\xi)=\V{U}_c'(\xi)$, also called the \textit{propagator mode}, is the corresponding eigenfunction. Because $\mathcal{L}$ is in general not self-adjoint, the eigenfunction $\V{W}^\dagger(\xi)$ of the adjoint operator $\mathcal{L}^\dagger$ to eigenvalue zero, the so-called \textit{response function}, is not
identical to $\V{W}(\xi)$. Expanding \eq{eq:RDA} with $\V{u}_0= \V{U}_c(\xi)+ \e \tilde{\V{u}}$ up
to $\mathcal O(\e)$ yields the PDE $\partial_t \tilde{\V{u}} = \mathcal{L} \tilde{\V{u}}+\V{v}(\xi+c_0 t)\cdot \nabla \V{U}_c$. Its solution $\tilde{\V{u}}$
can be expressed in terms of eigenfunctions $\V{w}_i$ of $\mathcal{L}$ as $\tilde{\V{u}}=\sum_i a_i(t) \V{w}_i(\xi)$ with expansion coefficients $a_i \sim \int_{t_0}^t\mathrm{d}\tilde{t} e^{\lambda_i (t-\tilde{t})} b(\tilde{t})$ and $b$ a functional of $\mathbb{D}\, \bracket{\V{v}\cdot\nabla}\V{u}_0$ involving eigenfunctions
of $\mathcal{L}^\dagger$; for details see Supplemental Material in Ref.~\cite{Loeber2014PRL}. By  multiple scale theory for small perturbations of the order $\e$ \cite{purwins2005,Biktashev2009,Loeber2014PRL},
the following equation of motion (EOM) for the position $\phi(t)$ of the TW in the presence of a boundary-induced advection term is obtained
\begin{align} \label{eq:eom}
 \dot{\phi}(t)=\!c_0-\!\frac{1}{K_c}\!\intl{-\infty}{\infty}{\xi}\V{W}^\dagger(\xi)^T\mathbb{D}\frac{Q'(\xi+\phi(t))}{Q(\xi+\phi(t))} \V{U}_c'(\xi),
\end{align}
with constant $K_c=\int_{-\infty}^{\infty}\mathrm{d}\xi\, \V{W}^\dagger(\xi)^T\,\V{U}_c'(\xi)$ and initial condition $\phi(t_0)=\phi_0$. For monotonically decreasing front
solutions, we define its position $\phi$ as the point of steepest slope, while for pulse solutions it is the point of maximum amplitude of an arbitrary component. The EOM \eqref{eq:eom} only takes into account the contribution of the perturbation $\mathbb{D}\, \bracket{\V{v}\cdot\nabla}\V{u}_0$ projected on the response function $\V{W}^\dagger(\xi)^T$ affecting the TW's position. Such EOM must be seen as the first two terms of an asymptotic series \cite{Bender1978}.

Since the integrand in \eq{eq:eom} does not explicitly depend on time, the mean time $T_c$ the TW needs to travel one period $L$ is given by
\begin{align}
 T_c =&\, \intl{0}{L}{\phi} \frac{1}{c_0- \Theta (\phi)},
 \intertext{and thus the average propagation velocity $c$ reads}
 c=&\, \frac{L}{T_c}= L \Big/ \intl{0}{L}{\phi} \frac{1}{c_0- \Theta (\phi)}, \label{eq:theo_c}
\end{align}
with substitute $\Theta (\phi) = \int_{-\infty}^{\infty}\mathrm{d}\xi\,\V{W}^{\dagger T} \mathbb{D} \frac{Q'(\xi+\phi)}{Q(\xi+\phi)} \V{U}_c'/K_c$.

\section{Schl\"ogl model} \label{sec:numerics}

\noindent In what follows, we limit our consideration to a single component system, $\V{u}=u$, with bistable reaction kinetics. The associated RD equation reads
\begin{align} \label{eq:Schlogl}
 \partial_t u =\,D_u \Delta u -u\bracket{u-a}\bracket{u-1},\quad 0<a<1
\end{align}
in dimensionless form. The parameter $a$ is related to the local excitation threshold of the medium while the two stable states are given by  $u=0$ and $u=1$, respectively. This model was introduced by Zeldovich and Frank-Kamenetskii in the modeling of flame propagation in 1938 \cite{Zeldovich1938} and then applied by Schl\"ogl to the description of a first-order non-equilibrium phase transition \cite{Schlogl1972}. Traveling front solutions to \eq{eq:Schlogl} obey the Dirichlet boundary conditions $\lim_{x\to -\infty} u(\V{r},t)=\,1$ and  $\lim_{x\to \infty} u(\V{r},t)=\,0$, respectively, and fulfill the condition $\lim_{\xi\to \pm\infty} \partial_\xi^n u = 0,\,\forall n \geq 1$.

In channels with non-modulated cross-section $Q(x)=\mathrm{const}$, \eq{eq:Schlogl} possesses a stable TF solution whose profile 
\begin{align}
 u(\V{r},t)=\,U_c(\xi)=\,\frac{1}{1+e^{\xi/\sqrt{2\,D_u}}},
\end{align}
and the corresponding propagation velocity
\begin{align} \label{eq:c0}
 c_0 = \sqrt{\frac{D_u}{2}}\bracket{1-2\,a},
\end{align}
are known analytically. The width of the traveling front
\begin{align}
 l= \sqrt{32\,D_u},
\end{align}
defines the intrinsic length scale \cite{Schlogl1972}. Noteworthy, the front velocity depends on the excitation thres\-hold $a$ while the front profile and, consequently, the front width are independent of $a$. This is a peculiarity of the Schl\"ogl model. We emphasize that the front width $ l$ depends solely on the diffusion constant $D_u$ in our scaling. Therefore, we can adjust the latter by means of the diffusion coefficient in the simulations. Furthermore, one can prove that the response function $W^\dagger(\xi)$ reads
\begin{align}
 W^\dagger(\xi)=\,e^{c_0\,\xi/D_u} U_c'(\xi),
\end{align}
and thus the constant $K_c$ is given by
\begin{align}
 K_c = \intl{-\infty}{\infty}{\xi} W^\dagger\,U_c'=\,\frac{\pi}{3}\sqrt{\frac{2}{D_u}}\frac{a\bracket{1-a}\bracket{1-2a}}{\sin\bracket{ 2 a \pi}}.
\end{align}
For the profiles of the perpendicular side-walls we chose a sinusoidally modulated boundary function
\begin{align} \label{eq:wall}
 \omega_+(x)\!=\!
 \begin{cases}
 1\! &\mbox{, for}\,x<0, \\ \frac{1}{2}\left[1+\delta+(1-\delta)\cos\bracket{\!\frac{2\pi x}{L}\!}\right]\!\!\!\!\!\!&\mbox{, for}\, x\geq 0,
\end{cases}
\end{align}
for the upper wall and set the lower boundary to $\omega_-(x)=0$. This setup is equivalent to study a reflection symmetric channel with twice the width, $\omega_\pm(x)=\pm\omega(x)$. The chosen boundary profile can be seen as the first harmonic of a Fourier series of a complicated periodic boundary profile.

Next, we compare our analytic estimate for the average front velocity $c$, \eq{eq:theo_c}, with numerical results. Therefore, \eq{eq:Schlogl} supplemented with the Neumann BC \eq{eq:bc} is solved numerically using finite element method (FEM) \cite{FreeFem2,*FreeFem}. In our simulations, the front is initialized with $U_c(x-x_\mathrm{start})$ at $x_\mathrm{start}=\mbox{min}(-4\,l, -L)$ and simulated until it reaches $x_\mathrm{end}= \mbox{max}(10\, L, \mbox{ceil}(10\,l /L) L)$. The data for the average front velocity $c$ are determined from a linear fit to a position vs. time plot after subtracting transients.

\begin{figure}[t]
  \centering
  \includegraphics[width=0.95\linewidth]{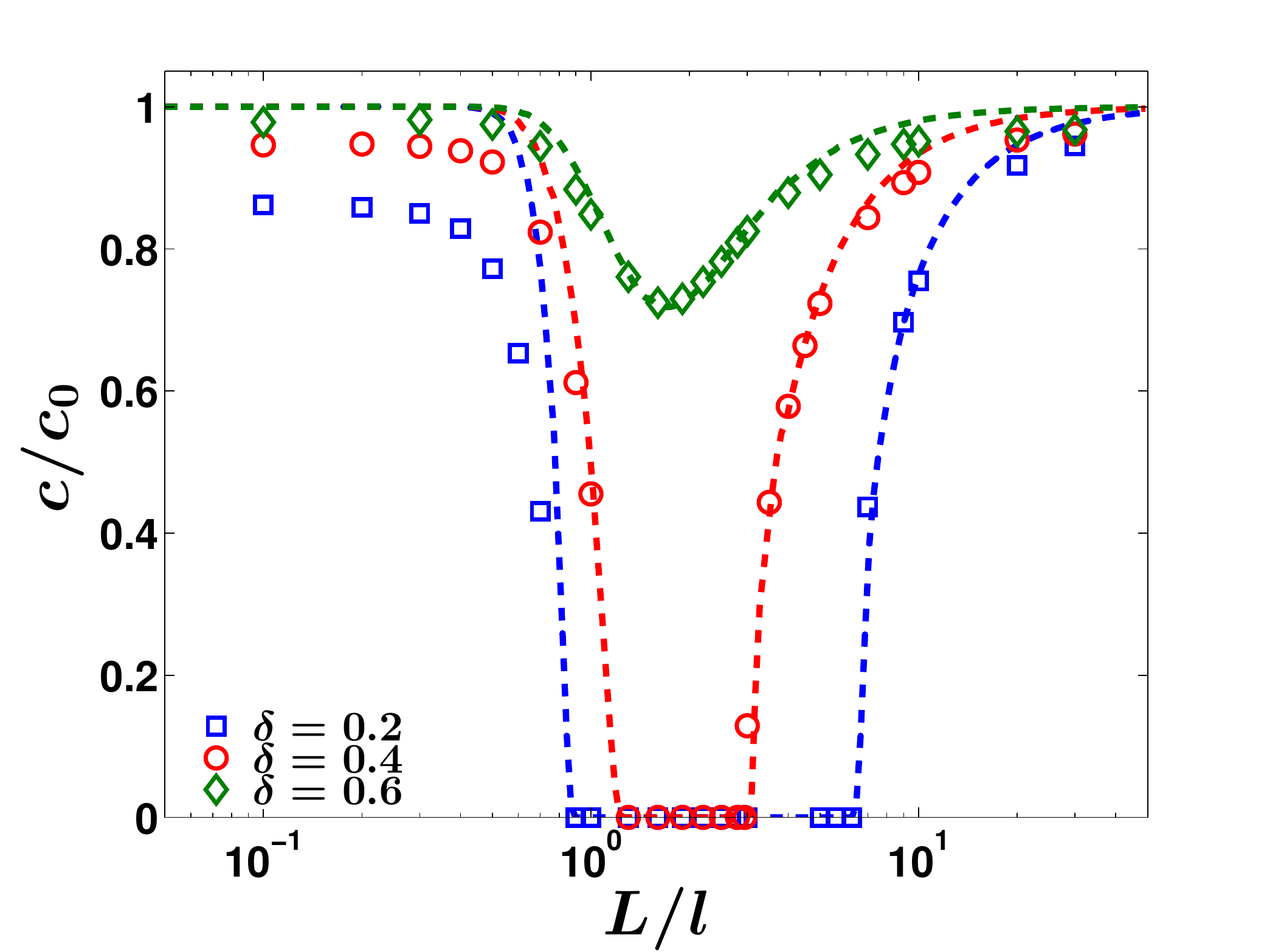}
  \caption{(Color online) Average front velocity $c$ in units of the propagation velocity $c_0$ as a function of the ratio of period length $L$ to front width $l$ for a sinusoidally modulated channel; see \figref{fig:Fig1}. The projection method (lines), \eq{eq:theo_c}, yields excellent agreement with the numerical results (markers); particularly, it reproduces the interval of propagation failure $c=0$ for intermediate values of $L/l$, however, it fails for small ratios $L/l \ll 1$. The remaining parameter values are set to $L=5$ and $a=0.4$.}
  \label{fig:Fig2}
\end{figure}

In \figref{fig:Fig2}, we depict the average front velocity $c$ in units of $c_0$ as a function of the ratio of period length $L$ to front width $l$. One observes a nonlinear dependence of $c$ on the ratio $L/l$: If the spatial period is much larger than the intrinsic front width, $L\gg l$, the front velocity equals $c_0$. In this limit, the front is well approximated by an iso-concentration line and one can assume that its velocity instantaneously adapts when traveling through the corrugated channel. Then, the average front velocity is correctly predicted by the harmonic mean velocity \cite{Loeber2012PRE}
\begin{align}
 c_\mathrm{harm}= L \Big/ \int_{0}^{L} \frac{\mathrm{d}x}{c_0+D_u Q'(x)/Q(x)},
\end{align}
which tends to $c_0$ for $L/l \to \infty$. With decreasing ratio $L/l$, i.e., either increasing the diffusion constant $D_u$ or decreasing the period length $L$, the average propagation velocity lessens until it attains its minimum value. Decreasing $L$ further, $L \lesssim l$, the value of $c$ grows and finally saturates at a value smaller than $c_0$.

It turns out that both the minimum value of $c$ and the saturation value depend crucially on the bottleneck width $\delta$. In general, we find that the average front velocity diminishes with shrinking bottleneck width $\delta$ for a given ratio $L/l$. In particular, we identify a finite interval of $L/l$ values where \textit{propagation failure} occurs, i.e., the initially traveling front becomes quenched \cite{Xin1995} and $c$ goes to zero. One observes that the lower bound of $L/l$ values, where the propagation failure interval begins, shrinks with decreasing bottleneck width $\delta$ while the upper bound, where the propagation failure interval ends, becomes larger for smaller bottlenecks. Consequently, the width of the propagation failure interval grows with decreasing value of $\delta$. Lowering the excitation threshold while
keeping the channel parameters $L$ and $\delta$ constant facilitates the traveling front to transit through the corrugated media and thus the interval of propagation failure disappears for $a\to 0$ (not shown explicitly).

Additionally, we compare our numerical results (markers) with the analytical prediction (lines), \eq{eq:theo_c}, in \figref{fig:Fig2}. Noteworthy, the analytic result matches excellent with the numerics for all bottleneck widths $\delta$. Moreover, it reproduces the interval of propagation failure for intermediate values of $L/l$, however, it fails for small ratios $L/l \ll 1$.

\begin{figure}[t]
  \centering
  \includegraphics[width=0.95\linewidth]{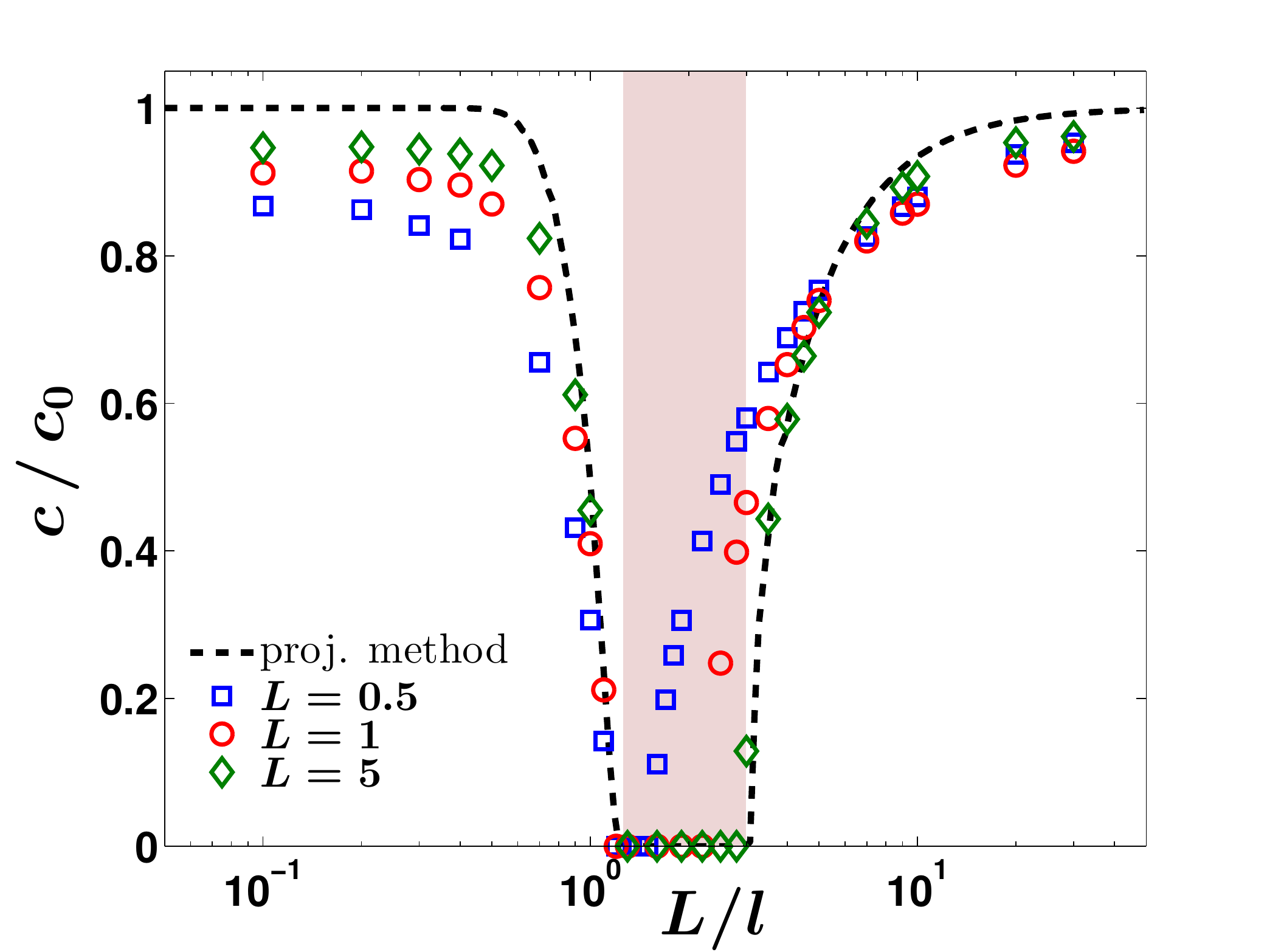}
  \caption{(Color online) Numerical results for propagation velocity $c$ versus $L/l$ for various period lengths $L=0.5$ (squares), $L=1$ (circles), and $L=5$ (diamonds). The corresponding values for the expansion parameter $\e$ are $\e=1.2,0.6$, and $\e=0.12$. The analytical prediction based on \eq{eq:theo_c} is represented by the dashed line and the predicted interval of propagation failure is indicated by the light red block. The remaining parameter values are set to $a=0.4$ and $\delta=0.4$.}
  \label{fig:Fig3}
\end{figure}

In \figref{fig:Fig3}, we illustrate a peculiarity of the projection method. The EOM for a TW under the boundary-induced perturbation \eq{eq:eom} is determined by the convolution integral $\Theta(\phi)$ with kernel $W^\dagger(\xi) U_c'(\xi)= e^{c_0 \xi /D_u} U_c'(\xi)^2$. The latter consists of the response function $W^\dagger(\xi)$ and the Goldstone mode $U_c'$ both being localized around $\xi=0$ and decay exponentially to zero if $ |\xi| > l$ for traveling fronts. On the other hand, the advection field $\V{v}(x)$ changes periodically with period $L$. Therefore, the dynamics of the front position $\dot{\phi}(t)$ and, consequently, the analytic result for the average front velocity $c$ depend solely on the ratio of $L$ to $l$ for a given set of $\delta$ and $a$ values. This is shown in \figref{fig:Fig3} where we present the propagation velocity $c/c_0$ for various period lengths, viz., $L=0.5$ (squares), $L=1$ (circles), and $L=5$ (diamonds). The value of $D_u$ is 
adjusted accordingly. It turns out that the analytic prediction based on project method \eq{eq:theo_c} agrees well with the numerics for $L=5$. With decreasing period length the range of applicability diminishes. This is in compliance with the assumptions made to derive \eq{eq:theo_c}: Both the asymptotic perturbation analysis, \secref{subsec:asym}, and the multiple scale analysis, \secref{subsec:proj}, require that the channel's cross-section changing rate $\mathrm{max}(|Q'(x)|) \propto \e$ is small. According to \eq{eq:epsilon}, the value of $\e$ is inversely proportional to the period length $L$ and thus the analytical prediction fails e.g. for $L=0.5$ ($\e=1.2$).

Moreover, we observe that the range of $L/l$ values where propagation failure occurs shrinks with decreasing period length $L$. Remarkably, the lower bound seems to be independent of $L$.

\subsection{Propagation failure -- eikonal approach} \label{subsec:eikonal}
\noindent Next, we present a qualitative explanation for the appearance of propagation failure for $L \gg l$. If the intrinsic front width is much shorter than the spatial period, any front propagating through the channel geometry can be well approximated by one time-dependent curve
$\grb{\gamma}(s,t)$ tracing out a chosen iso-concentration line parametrized
by $s$. A plane front $\grb{\gamma}=(c_0\,t, y)^T$ has constant velocity at all points in the forward, normal direction. If the plane geometry of the wave is distorted, here, due to the requirement that the edge of the wave front must propagate so that the TW always meets the boundary orthogonally, the normal velocity varies locally across the front.

When a front attempts to turn around a curved boundary, the entire front becomes curved and is well described by a circular arc with radius $r_c$ touching orthogonally the boundary $\omega_+(x)$. The associated curvature at $x*$ is given by
\begin{align} \label{eq:curv}
 \kappa(x*)=\,\frac{1}{r_c}=\, \frac{1}{\omega_+(x*)}\frac{\omega_+'(x*)}{\sqrt{1+\omega_+'(x*)^2}}\simeq \frac{\omega_+'(x*)}{\omega_+(x*)}.
\end{align}
Comparing \eqs{eq:eom} and \ref{eq:curv}, one notices that the EOM for TWs resembles the linear eikonal equation \cite{Keener1986,Dierckx2011}
\begin{align} \label{eq:eikonal}
 \dot{\phi}(t)\simeq c_0-D_u\,\kappa(\phi), 
\end{align}
in the limit $l\ll L$ and $W^\dagger(\xi) U_c'(\xi)\to \delta(\xi)$, respectively. 
According to \eq{eq:eikonal}, standing fronts exist and thus propagation failure occurs if the local curvature is equal to $\kappa(\phi)=c_0/D_u$.
Grindrod et. al \cite{Grindrod1991} demonstrated that stationary circular TW solutions are stable against deformations if the stability condition
\begin{align} \label{eq:stabcrit}
 \omega_+''(x*) \notin \left[0 , \omega_+'(x*)^2 \Big/ \bracket{\omega_+(x*) (1+\omega_+'(x*)^2)^2} \right]
\end{align}
holds at any $x* \in [0,L]$.

\begin{figure}[t]
  \centering
  \includegraphics[width=0.95\linewidth]{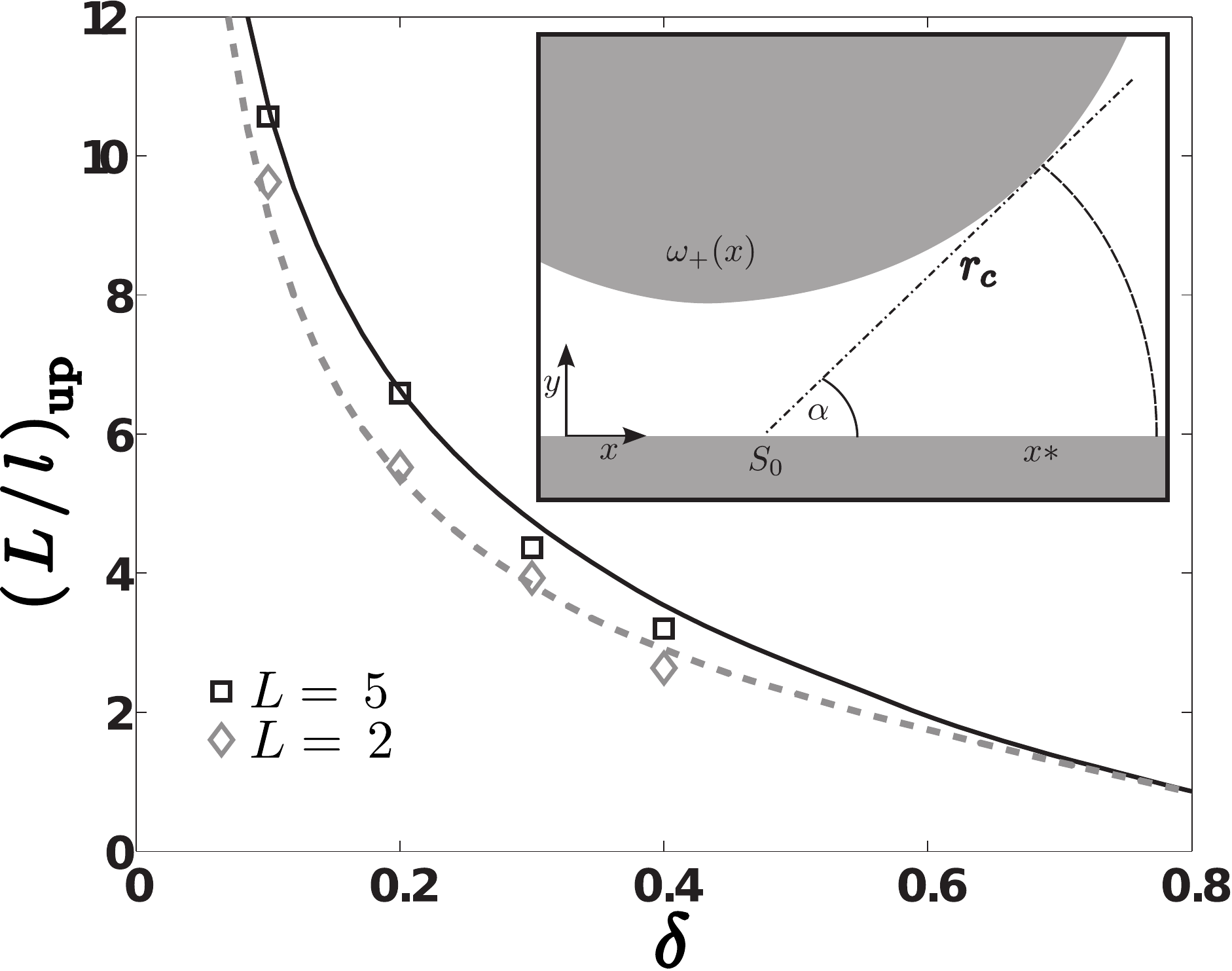}
  \caption{(Color online) Dependence of the upper bound $\bracket{L/l}_\mathrm{up}$ on the bottleneck width $\delta$. The eikonal approach ($L=2$: dashed line, $L=5$: solid line), \eqs{eq:eikonal}-\eqref{eq:stabcrit}, yields excellent agreement with the numerical results ($L=2$: diamonds, $L=5$: squares) for narrow channels $\delta \ll 1$ but fails for channels with wide openings. In simulations no propagation failure was found for $\delta \gtrsim 0.5$. The numerical errors are of the size of the markers. Inset: Sketch of a circular iso-concentration line with radius $r_c$ touching orthogonally the top boundary $\omega_+(x)$ at $x*$.
  }
  \label{fig:Fig4}
\end{figure}

In order to determine the upper bound of $L/l$, first one has to find the roots of $0=c_0-D_u \kappa(x*)$ and secondly has to check if the stability condition \eq{eq:stabcrit} is not satisfied at $x*$. The dependence of the upper bound $\bracket{L/l}_\mathrm{up}$ on the bottleneck width $\delta$ is depicted in \figref{fig:Fig4}. Moreover, we compare the analytic predictions based on the linear eikonal approach (lines) with the numerically obtained values for $\bracket{L/l}_\mathrm{up}$ using FEM simulations (markers). Obviously, the agreement is excellent for small bottlenecks but fails for large values of $\delta$. In channels with wide openings $\delta \to 1$ traveling waves have to curve only close to edge of the wave front. Consequently, the iso-concentration lines $\grb{\gamma}(s,t)$ are almost planar, i.e., the front travels with almost constant velocity in $x$-direction, and propagating failure does not occur.

\section{Diffusion limited regime -- confined Brownian motion} \label{sec:diff}

\noindent In Figs. \ref{fig:Fig2} and \ref{fig:Fig3}, we have shown that the average propagation velocity saturates at a value which is smaller than $c_0$ if the front width is much larger than the period of the channel, $L / l \to 0$. We emphasize that the projection method fails in this limit and predicts $c=c_0$ regardless of the value of $L$ and $\delta$. In contrast, the numerics show that the saturation value lessens with decreasing bottleneck width $\delta$, cf. \figref{fig:Fig2}, and with shrinking period $L$, see \figref{fig:Fig3}; in a word, with growing value of $\e=(1-\delta)/L$. If the intrinsic width $l=\sqrt{32 D_u}$ is much larger than the periodicity of the channel $L$, the front is extended over many periods and boundary interactions play a subordinate role. Then, the diffusion of reactants in propagation direction under spatially confined conditions is the predominant process for wave propagation and the problem can be approximated by an effective one-
dimensional description introducing an effective diffusion constant $D_\mathrm{eff}$
\begin{align} \label{eq:RDeff}
 \partial_t \V{u}(\V{r},t) = D_\mathrm{eff} \partial_x^2 \V{u} + \V{R}(\V{u}).
\end{align}
Experimental \cite{Verkman2002,Corma1997} and theoretical studies \cite{Reguera2006,Burada2008} on particle transport in micro-domains with obstacles \cite{Dagdug2012a,Martens2012} and/or small openings revealed that Brownian motion in such systems exhibits non-intuitive features like a significant suppression of particle diffusivity \cite{Cohen2006,Martens2013,Keyer2014,*Keyer2014arxiv}. Numerous research activities in this topic led to the development of an approximate description of the diffusion problem -- the \textit{Fick-Jacobs approach} \cite{Jacobs,Zwanzig1992}. The latter provides a powerful tool to describe particle transport through corrugated channel geometries and its accuracy has been intensively studied for diffusing particles in two- \citep{Burada2008,Burada2009_CPC} and three-dimensional channels \citep{Dagdug2011,*Berezhkovskii2007}. The Fick-Jacobs approach predicts that the effective diffusion constant in longitudinal direction is solely determined by the
variation of the cross-section $Q(x)$ and can be calculated according to the Lifson-Jackson formula \cite{Lifson1962}
\begin{align}
 D_\mathrm{eff}^\mathrm{FJ} = \frac{D_u}{\av{Q(x)}_x\,\av{1/Q(x)}_x}
\end{align}
with period average $\av{\cdot}_x=L^{-1}\int_{0}^{L} \cdot \,\mathrm{d}x$. For the exemplarily chosen channel geometry, \eq{eq:wall}, the value of $D_\mathrm{eff}$ is given by
\begin{align} \label{eq:DeffFJ}
 D_\mathrm{eff}^\mathrm{FJ}=\,D_u \,\frac{2\sqrt{\delta}}{1+\delta}.
\end{align}
Similar to the derivation of the reaction-diffusion-advection equation for $\V{u}_0$, see \secref{subsec:asym}, the Fick-Jacobs approach is valid solely for weakly modulated channel geometries, i.e., $\mathrm{max}|Q'(x)| \propto \e \ll 1$. For moderate to strong corrugated boundaries, higher order correction terms have to be considered \cite{Martens2011}, yielding
\begin{align}\label{eq:Deffeps}
 D_\mathrm{eff}^\e=\,D_u \,\frac{4\,L\,\sqrt{\delta}}{\pi\bracket{1-\delta^2}}\,\asinh\bracket{\frac{\pi\bracket{1-\delta}}{2 L}}.
\end{align}
For the studied Schl\"ogl model, the average propagation velocity in units of the free velocity might be approximated well by $c/c_0\simeq\sqrt{D_\mathrm{eff}^{\mathrm{FJ},\e}/D_u}$, cf. \eq{eq:c0}.

\begin{figure}[t]
  \centering
  \includegraphics[width=0.95\linewidth]{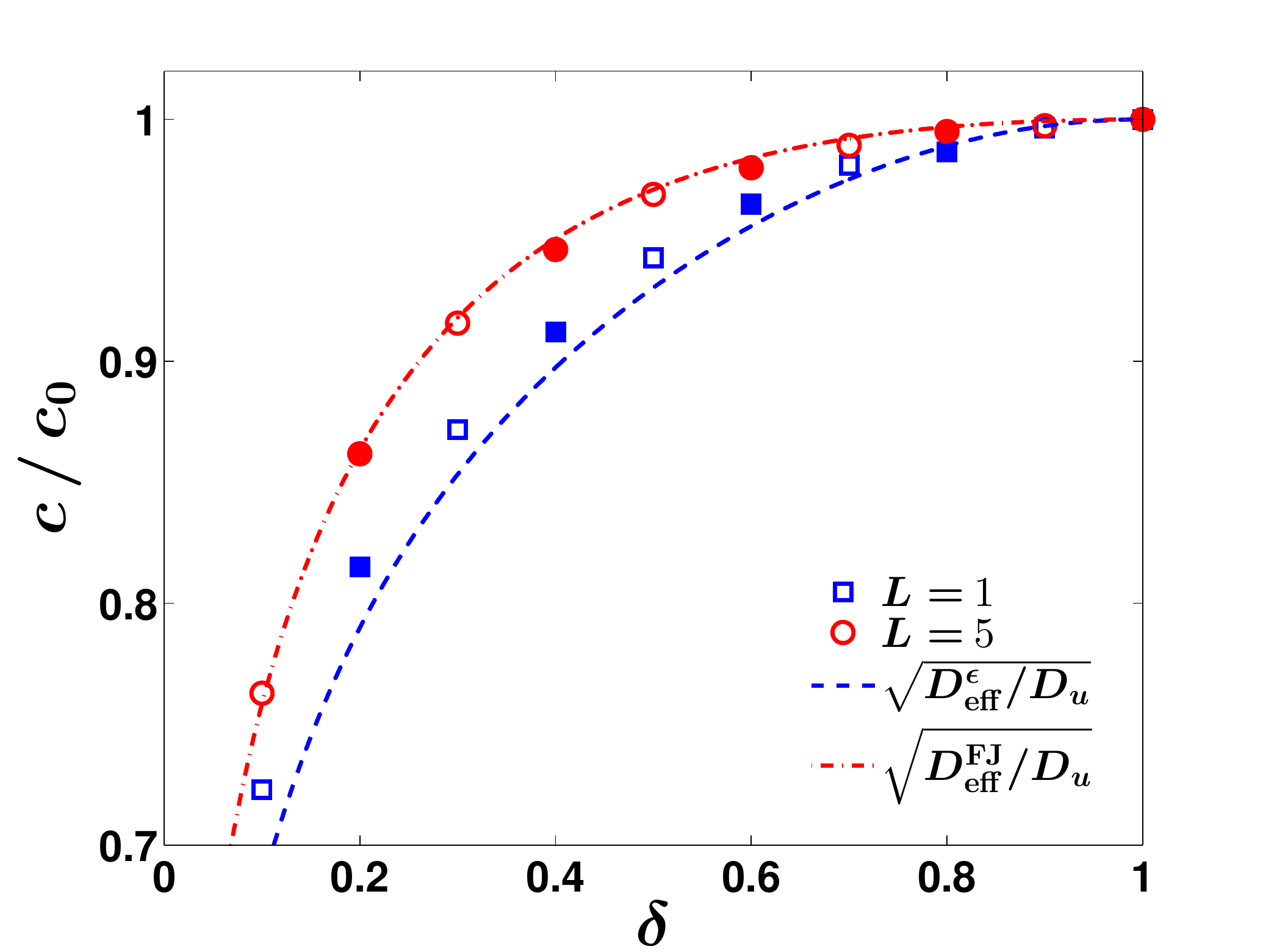}
  \caption{(Color online) Average front velocity $c$ versus bottleneck width $\delta$ for different values of $a=0.2$ (blank markers) and $a=0.4$ (filled markers). The ratio $L/l$ is set to $L/l=0.1$ and the diffusion coefficient $D_u$ is adjusted accordingly, viz., $D_u=3.125$ ($L=1$) and $D_u=78.125$ ($L=5$).}
  \label{fig:Fig5}
\end{figure}

In \figref{fig:Fig5}, we present the impact of the bottleneck width $\delta$ on the propagation velocity $c$ in units of $c_0$ for two different period lengths, viz., $L=1$ and $L=5$. In the numerics, the ratio $L/l$ is set to $L/l=0.1$ and the diffusion coefficient $D_u$ is adjusted accordingly. One notices that the analytic estimate using $D_\mathrm{eff}^\mathrm{FJ}$, \eq{eq:DeffFJ}, agrees excellently with our simulation results for large period lengths $L=5$ (circles) and small values of $\e$, $\e \in [0,0.2]$, respectively. For smaller periods, $L=1$ (squares), higher order corrections to the effective diffusion coefficient \eq{eq:Deffeps} are necessary in order to ensure a good agreement between numerics and analytics. The corresponding value for the expansion parameter $\e$, \eq{eq:epsilon}, ranges from zero to unity. Remarkably, the saturation value for the front velocity in units of its free value $c_0$ is independent of the value for the excitation threshold $a$
and thus solely determined by the spatial variations of the channel cross-section $Q(x)$; to sum up, $c\simeq\sqrt{0.5 D_\mathrm{eff}(Q(x))}(1-2 a)$ for $l \gg L$.

Interestingly, in the limit $L \ll l$ the propagation of traveling fronts through channels with spatially modulated cross-sections can be well treated by a one-dimensional RD equation \eq{eq:RDeff}. Within the latter, the impact of the spatial variations on the reactants' microscopic dynamics is neglected. 
However, the influence of boundary modulation on diffusive transport of material is hidden in an artificially introduced effective diffusion coefficient $D_\mathrm{eff}$. To estimate the value of $D_\mathrm{eff}$ detailed information about the shape of the cross-section is needed \cite{Brenner}.


\section{Conclusion} \label{sec:conc}

\noindent We have investigated the propagation of reaction-diffusion waves confined to a channel with walls impermeable to diffusion. In propagation direction the channel's cross-section changes periodically on the length scale $L$. For weak modulations of the channel's cross-section the space-dependent no-flux boundary conditions can be mapped on a \textit{boundary-induced advection term}. The latter is proportional to the spatial variation of the cross-section $Q(x)$. Using projection method, we derive an equation of motion for the position of a traveling wave as function of time in the presence of the boundary-induced advection term. From the latter, we obtain an analytical expression for the average propagation velocity $c$ of the wave traveling through periodically modulated channels. 

Exemplary, we study the impact of a sinusoidally modulated cross-section on the propagation of traveling front solutions in a one-component Schl\"ogl model. It turns out that the propagation velocity exhibits a nonlinear dependence on the ratio of the spatial period $L$ to the intrinsic width of the front $l$: If the period is much larger than the intrinsic width, $L\gg l$, a Schl\"ogl front travels at the harmonic mean velocity which tends to the value for non-modulated channels $c_0$. With decreasing ratio $L/l$ the average propagation velocity lessens, attains its minimum value, and starts to grow again until it finally saturate at a value below the velocity in the unperturbed channel for $L\ll l$. 

Beyond a critical bottleneck width, \textit{propagation failure} occurs, i.e., the initially traveling front becomes quenched inside the corrugated channel and hence the minimal propagation velocity vanishes identically. With decreasing bottleneck width, the lower and upper bound for propagation failure shift to smaller and larger values for $L/l$, respectively. While the shift is almost independent of $L$ for the lower bound, the upper bound grows with $L$. Moreover, we have demonstrated that the existence of propagation failure and, in particular, the dependence of the upper bound of $L/l$ can be completely understood based on the linear eikonal approach.

In the case of very small periods, $L/l\ll 1$, the front velocity is determined solely by the shape of the cross-section. In this limit, front propagation is dominated by the diffusive motion of the reactants in spatial confinement. The spatial dependent no-flux boundary conditions on the reactants translate into a one-dimensional reaction-diffusion system with an effective diffusion coefficient $D_\mathrm{eff}$ as it is demonstrated by the excellent agreement with simulation results. Thereby, the influence of the spatial confinement on the microscopic dynamics is hidden in the value of $D_\mathrm{eff}$ and Luther´s law is recovered.

Altogether, over a large range of spatial periods and bottleneck values, the analytical result for the averaged propagation velocity in the corrugated channel (including propagation failure) agrees remarkably well with numerical results obtained in FEM simulations. Since the applicability of our perturbation analysis is based on a small channel's cross-section changing rate, $\mathrm{max}(|Q'(x)|) \propto \e$, deviations from the analytical predictions are expected to grow for geometries with short-scale spatial modulations and narrow openings. Our results might be interesting for control purposes: In this case, a given protocol of movement for a traveling wave, $\phi(t)$, is realized by a space-dependent cross-section to be derived solving the integral equation \eq{eq:eom} for the unknown $Q(x)$, compare \cite{Loeber2014PRL}. For periodically varying cross-sections, $Q(x)=Q(x+L)$, accessible control parameters include the spatial period, the modulation amplitude, and the bottleneck width. Results in this 
direction will be presented in a forthcoming paper.

\acknowledgments
\noindent We acknowledge support by the DFG through SFB 910 (S. M and H. E.) and GRK 1558 (J. L.).


%

\end{document}